\documentclass[12pt]{iopart}
\usepackage{epsfig}
\def\ib#1,#2,#3{       {\it ibid.\/ }{\bf #1} (19#2) #3}
\def\ap#1,#2,#3{       {\it Ann.~Phys.~(NY)\/ }{\bf #1} (19#2) #3}
\def\ijmp#1,#2,#3{     {\it Int.\ J.~Mod.\ Phys.\/ } {\bf A#1} (19#2) #3}
\def\mpl#1,#2,#3 {     {\it Mod.~Phys.~Lett.\/ } {\bf A#1} (19#2) #3}
\def\npb#1,#2,#3{       {\it Nucl.\ Phys.\/ }{\bf B#1} (19#2) #3}
\def\npps#1,#2,#3{     {\it Nucl.\ Phys.~B (Proc.~Suppl.)\/ }{\bf B#1}
                             (19#2) #3}
\def\plb#1,#2,#3{      {\it Phys.\ Lett.\/ }{\bf B#1} (19#2) #3}
\def\pr#1,#2,#3{       {\it Phys.\ Rev.\/ }{\bf #1} (19#2) #3}
\def\prd#1,#2,#3{      {\it Phys.\ Rev.\/ }{\bf D#1} (19#2) #3}
\def\prep#1,#2,#3{     {\it Phys.\ Rep.\/ }{\bf #1} (19#2) #3}
\def\prl#1,#2,#3{      {\it Phys.\ Rev.\ Lett.\/ }{\bf #1} (19#2) #3}
\def\pro#1,#2,#3{      {\it Prog.~Theor.\ Phys.\/ }{\bf #1} (19#2) #3}
\def\rmp#1,#2,#3{      {\it Rev.~Mod.~Phys.\/ }{\bf #1} (19#2) #3}
\def\sp#1,#2,#3{       {\it Sov.~Phys.~Usp.\/ }{\bf #1} (19#2) #3}
\def\zpc#1,#2,#3{      {\it Z.~Phys.\/ }{\bf C#1} (19#2) #3}
\def\appb#1,#2,#3{     {\it Acta Phys.\ Polon.\/ }{\bf B#1} (19#2) #3}

\def\subfigureA#1{\leavevmode\hbox{#1}}

\begin{document}
%
\hfill IFT/98-30

\title[Contact Interactions with Polarized Beams at HERA]
  {Contact Interactions with Polarized Beams  at HERA}

\author{J.Kalinowski $^{1}$, H. Spiesberger $^{2}$, J.M.Virey $^{3}$} 

\address{$^{1}$ Instytut Fizyki Teoretycznej, Uniwersytet Warszawski,
  PL-00681 Warszawa, Poland} 
\address{$^{2}$ Institut f\"ur Physik, Johannes-Gutenberg-Universit\"at,
  D-55099 Mainz, Germany} 
\address{$^{3}$ Centre de Physique Th\'eorique, C.N.R.S. - Luminy, Case
  907, F-13288 Marseille Cedex 9, and Universit\'e de Provence,
  Marseille, France} 

\begin{abstract}
  The discovery potential of the HERA collider, with and without
  polarized beams, for electron-quark contact interactions in neutral
  current scattering is reviewed. The measurement of spin asymmetries in
  the polarized case could give crucial information on the chiral
  structure of new interactions.
\end{abstract}

\section{Introduction}
The study of contact interactions (CI) is a powerful way to search for
departures from the Standard Model and to parametrise new physics
phenomena. Phenomenologically at an energy scale much lower than the
characteristic scale of the underlying theory, very different new
physics contributions can give rise to very similar changes in
processes involving only SM particles.  The presence of new
interactions can then be written in terms of an effective CI
Lagrangian in a model-independent way.  In the context of HERA the
lepton-quark CI of the first generation fermions are the most
interesting since they can manifest themselves in $eq \rightarrow eq$
scattering. The most general $eeqq$ current-current effective
Lagrangian has the form \cite{ELP83}:
\begin{equation}\label{Leq}
{\cal L}_{eq}^{NC} = \sum_{q=u,d}\; \sum_{i,j=L,R} \eta^q_{ij}(\bar e_i 
\gamma_{\mu} e_i) 
(\bar q_j \gamma^{\mu} q_j) 
\end{equation}
with $\eta^q_{ij}= 4\pi \epsilon^q_{ij} /(\Lambda^q_{ij})^2$, where
$\Lambda^q_{ij}$ is the effective mass scale of the contact interaction.
The sign $\epsilon^q_{ij}$ characterises the nature of the interference
of each CI term with the Standard Model $\gamma$ and $Z$ exchange
amplitudes.  The subscripts $LL$, $RR$, $LR$ and $RL$ refer to the
chiral structure of the new interaction.  In a purely phenomenological
approach all $\eta$'s are unknown parameters, whereas they are predicted
in a given theoretical framework. For example, the structure of
$\eta^q_{ij}$ in a generic leptoquark model or in supersymmetry with
$R$-parity violation can be found in Ref~\cite{krsz}.

The overwhelming success of the SM suggests that the CI Lagrangian must
be $SU(2)_L\times U(1)$ symmetric. The symmetry implies that
$\eta^u_{RL}= \eta^d_{RL}$, relates $eq$ and $\nu q$ NC interactions,
and relates the difference $\eta^u_{LL}-\eta^d_{LL}$ to the CC $e\nu ud$
contact term.  The lepton-hadron universality of charged-current data
indicates that the latter is small. HERA will be most sensitive to CI
terms in measurements at large $Q^2$ which favours large $x$ where the
leading contribution comes from the valence $up$-quark. Therefore, for
simplicity, we will assume {\it u-d} universality, {\it i.e.}
$\eta^u_{ij}=\eta^d_{ij}$. As a result we are left with eight terms in
eq.~(\ref{Leq}) defined by various combinations of chiralities and
signs, which we will denote by $ij^{\epsilon}$, {\it i.e.}  $LR^+$,
$LL^-$ etc.

A global study of the $eq$ CI, based on the most relevant existing
experimental data, has recently been performed in Ref~\cite{Bargerct2}.
Stringent bounds of the order of $\Lambda \sim 10$ TeV for the
individual CI terms are found. However, when several terms of different
chiralities are involved simultaneously, cancellations occur and the
resulting bounds on $\Lambda$ are considerably weaker and of the order
of $3-4$ TeV \cite{Bargerct2}. Present bounds from HERA as well as from
other high-energy experiments are of the same order of magnitude and it
can therefore be expected that experiments at HERA will be able to
improve the limits with increasing luminosity. 

\section{Chiral structure of CI: the case for polarized beams}

In this note we want to emphasize that the measurement of spin
asymmetries, defined in the context of HERA with polarized lepton beams
($e^+$ and $e^-$) and/or with polarized lepton and proton beams, could
provide very important tools to disentangle the chiral structure of the
new interaction. To illustrate the point we will consider an example
with double spin asymmetries defined in eq.~(\ref{asym}) below.  The
details of a more complete phenomenological analysis of the experimental
signatures of CI from the measurements of cross sections and spin
asymmetries in the NC channel at HERA can be found in \cite{jmvhera}.
The analysis of \cite{jmvhera} has been motivated by the renewed
interest in the polarization option, considered already at an early
stage of the HERA project (see for example \cite{heraw}), and more
recently in \cite{DeRoeck}; we also refer to the report of the Working
Group 6 in these proceedings.

When we discuss polarized beams we split equally the expected total
integrated luminosity of 1 fb$^{-1}$ among various configurations of
beams and polarizations, {\it i.e.} for both lepton and proton beams
polarized, we assume a luminosity of 125 pb$^{-1}$ for $e^+p$ and
$e^-p$ with longitudinally polarized leptons ($\lambda_e=\pm$) and
protons ($\lambda_p=\pm$). As a result, the ``discovery potential''
(as far as the sensitivity to the scale $\Lambda$ is concerned) is not
significantly improved by running with polarized beams as compared to
the unpolarized case, see Table 1. There the 95\% CL limits for
$\Lambda$'s obtained from the analysis of unpolarized $e^+p$ and
$e^-p$ collisions (upper row) are compared to the ones obtained with
the help of polarized beams (lower row). In the latter case the
following double-spin asymmetries\footnote{The asymmetries defined in
  eq.~(\ref{asym}) are sensitive to the violation of parity, and are
  also interesting from the point of view of spin structure functions
  \cite{DeRoeck,agk}.}  have been used
\begin{equation}\label{asym}
A_{LL}^{PV} (e^-)\; =\; \frac{\sigma^{--}_- \, -\,
\sigma^{++}_-}{\sigma^{--}_- \,  
+\, \sigma^{++}_-}
\;\;\;\;\;\;\; \mbox{\rm and} \;\;\;\;\;\;\;
A_{LL}^{PV} (e^+)\; =\; \frac{\sigma^{--}_+ \, -\,
\sigma^{++}_+}{\sigma^{--}_+ \,  
+\, \sigma^{++}_+}\; 
\end{equation}
which are defined in terms of the polarized differential cross
sections $\sigma^{\lambda_e,\lambda_p}_l\equiv \mbox{d}
\sigma^{\lambda_e,\lambda_p}_l /\mbox{d} Q^2$ with $l=+$ for $e^+p$
and $l=-$ for $e^-p$. Exploiting other spin asymmetries defined in
Ref.~\cite{jmvhera} (see also \cite{Martyn91}) does not improve
significantly the limits.

\hspace{-0.8cm}
\begin{table}
\label{table1}
\begin{center}
\begin{tabular}{|c||c|c|c|c|c|c|c|c|}
\hline
$\Lambda$ (TeV)& $LL^+$ &$RR^+$ &$LR^+$ &$RL^+$ &$LL^-$ &$RR^-$ &$LR^-$ 
&$RL^-$\\
\hline
\hline
unpolarized & 6.3 & 6.1 & 6.2 & 6.0 & 5.5 & 5.3 & 5.2 & 5.0 \\
\hline
polarized &  5.5 & 6.0 & 5.3 & 6.0 & 5.4 & 5.8 & 5.1 & 5.8 \\
\hline
\end{tabular}
\end{center}
\caption{Limits on $\Lambda$ at 95\% CL for the unpolarized
and polarized cases. In the latter case the double-spin asymmetries of
eq.~(\ref{asym}) have been used to derive the limits.}
\end{table} 

\begin{figure}[htb]
\begin{tabular}[t]{c c}
\centerline{\subfigureA{\psfig{file={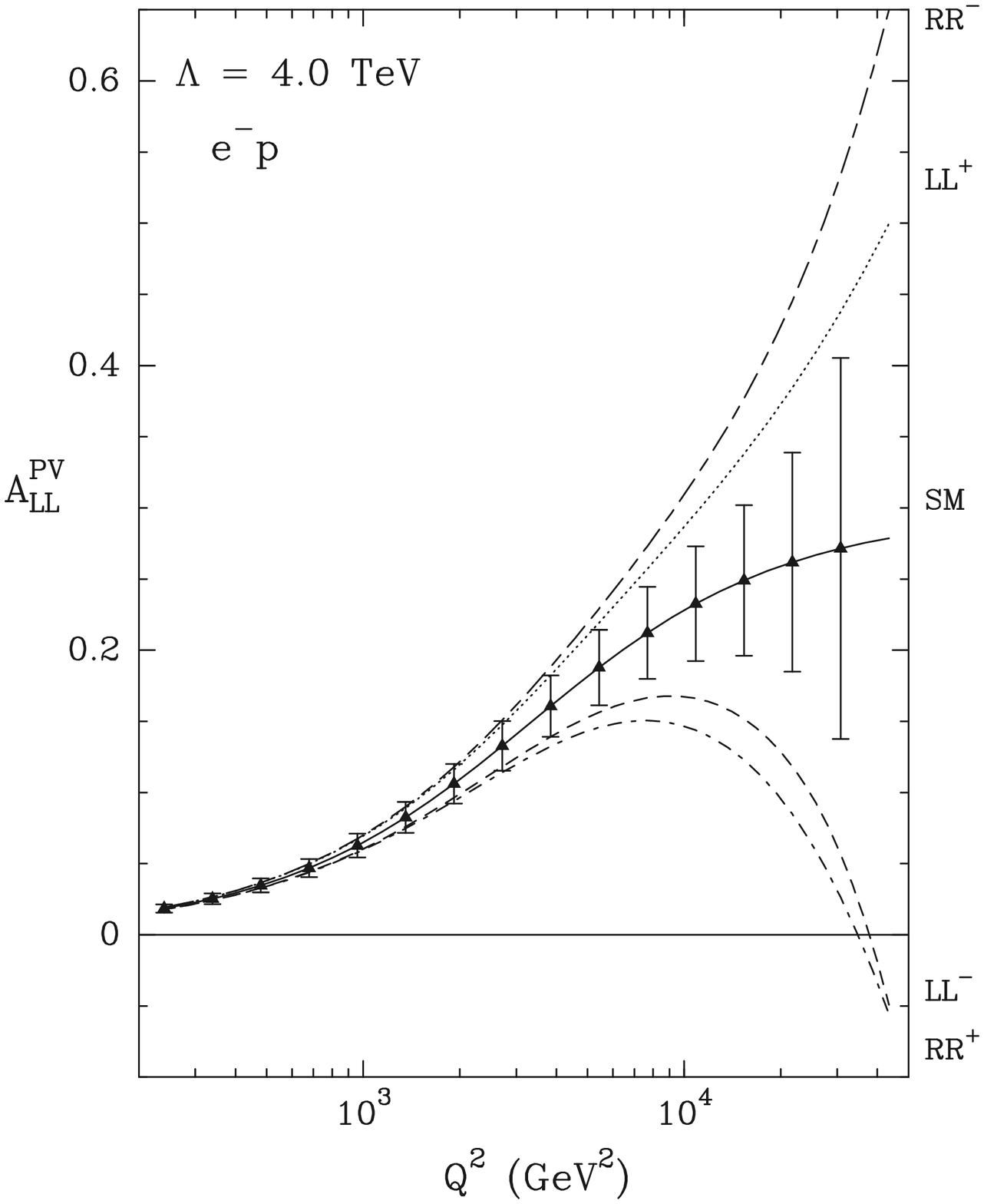},width=8truecm,height=10truecm}}
\subfigureA{\psfig{file={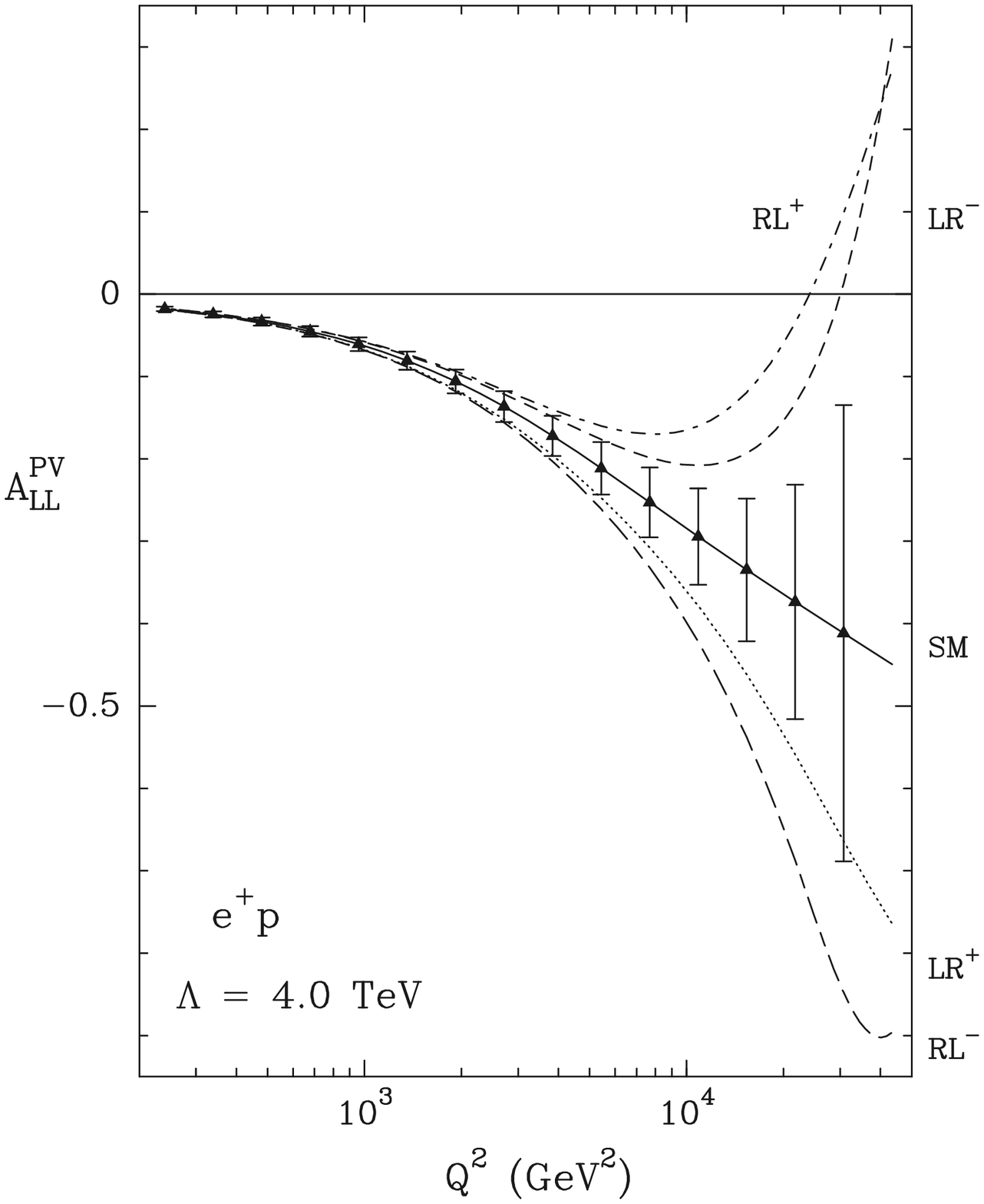},width=8truecm,height=10truecm}}}
\end{tabular} 
\vspace{-1cm}
\caption{\small 
  Spin asymmetries $A_{LL}^{PV} (e^-)$ and $A_{LL}^{PV} (e^+)$.  Solid
  lines correspond to the SM prediction; the expected errors are shown
  assuming a luminosity of 125 pb$^{-1}$ for each configuration of
  beam polarizations. Non-solid lines correspond to CI scenarios with
  $\Lambda = 4$ TeV and helicities as indicated.} 
\label{fig1}
\end{figure}

Once new physics effects are observed, polarized beams are very useful
since the $Q^2$ dependence of spin asymmetries contains additional
information which is sensitive to the chiral structure.  This is
exemplified in Fig.~\ref{fig1} where the $Q^2$ dependence of
$A_{LL}^{PV} (e^-)$ and $A_{LL}^{PV} (e^+)$, assuming $\Lambda=4$ TeV,
is drawn for several $ij^\epsilon$ CI terms. The observation of a
deviation from the SM in $e^-p$ will allow us to distinguish between
$LL^+/RR^-$ and $LL^-/RR^+$, and from $e^+p$ data between $LR^+/RL^-$
and $LR^-/RL^+$ contact interaction terms. Other spin asymmetries can
be exploited to reveal the anatomy of the chiral structure of contact
interactions \cite{jmvhera}.

\section{Conclusions}

The main conclusions are the following:\\
1) The HERA collider with an integrated luminosity of $L_{tot}=1$
fb$^{-1}$ will give strong bounds on the energy scale of a possible new
CI. For constructive interferences, the limit on $\Lambda$ is of the
order of $6\; TeV$, and for destructive interferences we get $\Lambda
\sim 5\; TeV$. The availability of polarized lepton and proton beams
will not increase significantly these bounds, except for destructive
interferences.  With only leptons polarized, the sensitivity is
strongly reduced.\\
2) The studies of spin (and charge) asymmetries can give unique
information on the chiral structure of the new interaction. The
availability of electron and positron beams is mandatory in order to
cover all the possible chiralities. \\
3) Since in the proton the valence $u$-quark distribution is dominant,
the measurements essentially constrain the presence of a new interaction
in the $eu$ sector. To constrain a new interaction in $ed$ sector,
protons could be replaced by neutrons, for example by using
electron-$He^3$ collisions, an option which is also under consideration
at HERA \cite{DeRoeck}.

\section{Acknowledgments} 
JK has been partially supported by the Polish Committee for Scientific 
Research Grant 2 P03B 030 14.


\section{References}
\begin{thebibliography}{99}
\bibitem{ELP83} Eichten E, Lane K and Peskin M 1983 \PRL {\bf 50} 811;\\    
R\"uckl R 1983 \PL {\bf B129} 363, 1984 \NP {\bf B234} 91

\bibitem{krsz} Kalinowski J, R\"uckl R, Spiesberger H and Zerwas P M
  1997 \ZP {\bf C74} 595 
  
\bibitem{Bargerct2} Barger V {\it et al} 1998 \PR {\bf D57} 391;
  Zeppenfeld D {\it et al} 1998 {\it preprint hep-ph/9810277}

\bibitem{jmvhera} Virey J M 1998 {\it preprint hep-ph/9809439} to be
  published in {\it Eur.\ Phys.\ J.\ } {\bf C}

\bibitem{heraw} Peccei R D {\it Ed} 1987 {\it Proc.\ 
  HERA Workshop};\\
  Buchm\"uller W and Ingelman G {\it Eds} 1991 {\it Proc.\ Physics at HERA}
  
\bibitem{DeRoeck} De Roeck A and Gehrmann T {\it Eds} 1997 {\it Proc.\ 
    Workshop on Physics with Polarized Protons at HERA}

\bibitem{agk} Anselmino M, Gambino P and Kalinowski J 1994 \ZP {\bf C64} 267

\bibitem{Martyn91} Martyn H U 1987 in {\it Proc.\ HERA Workshop 1987};\\
   Haberl P, Schrempp F and Martyn H U 1991 in {\it Proc.\ 
    Physics at HERA 1991}

\end{thebibliography}
\end{document}